\documentclass[12pt]{iopart}

\usepackage{iopams}
\usepackage{graphicx}
\usepackage{color}

\usepackage{textcomp}
\usepackage{bm}

\usepackage{cite}

\begin{document}
\title{Smooth double critical state theory for type-II superconductors}
\author{H. S. Ruiz}
\ead{hsruizr@unizar.es}
\address{Departamento de F\'{\i}sica de la Materia
Condensada--I.C.M.A., Universidad de Zaragoza--C.S.I.C., Mar\'{\i}a
de Luna 1, E-50018 Zaragoza, Spain}

\author{A. Bad\'{\i}a\,--\,Maj\'os}
\address{Departamento de F\'{\i}sica de la Materia
Condensada--I.C.M.A., Universidad de Zaragoza--C.S.I.C., Mar\'{\i}a
de Luna 1, E-50018 Zaragoza, Spain}
\date{\today}
\begin{abstract}

Several aspects of the general theory for the critical states of a vortex lattice and the magnetic flux dynamics in type-II superconductors are examined by a direct variational optimisation method and widespread physical principles. Our method allows to unify a number of conventional models describing the complex vortex configurations in the critical state regime. Special attention is given to the discussion of the relation between the flux-line cutting mechanism and the depinning threshold limitation. This is done by using a smooth double critical state concept which incorporates the so-called isotropic, elliptical, T and CT models as well-defined limits of our general treatment. 
Starting from different initial configurations for a superconducting slab in a 3D magnetic field, we show that the predictions of the theory range from the collapse to zero of transverse magnetic moments in the isotropic model, to nearly force free configurations in which paramagnetic values can arbitrarily increase with the applied field for magnetically anisotropic current voltage laws. Noteworthily, the differences between the several model predictions are minimal for the low applied field regime.

\end{abstract}
\maketitle


\section{Introduction}\label{Section_I}
%

%
The study of the critical state theory of a vortex lattice in type-II superconductors is a stimulating problem. It relates to a wide list of physical phenomena and also affects a number of practical applications. The original concept of a critical state dates back to the work by C. P. Bean~\cite{Bean_1962,Bean_1964,Bean_1970} who assumed that external magnetic field variations are opposed by the maximum current density in the material, i.e.: when non-vanishing $|{\bf J}|=J_{c}$. Physically, the driving force due to the currents circulating in the superconducting sample is balanced by the limiting pinning force acting on the vortex lattice so as to prevent destabilisation and the consequent propagation of dissipative states. It occurs that Bean's simplifying ansatz straightforwardly leads to predict the proper response of the sample provided the electrical current density vector $\textbf{J}$ is perpendicular to the local magnetic field vector $\textbf{B}$, i.e, $\textbf{J}(\textbf{r})=\textbf{J}_{c\perp}(\textbf{r})$. Just recall that magnetostatic forces are given by $\textbf{J}\times\textbf{B}$. However, 
unless for highly symmetric situations, $\textbf{J}$ does not necessarily satisfy the condition $\textbf{J}=\textbf{J}_{\perp}$. Therefore, the stronger limitation of Bean's model is that one can just apply it to vortex lattices composed by parallel flux lines perpendicular to the current flow.
On the other hand, rotations of $\textbf{B}$ can lead to entanglement and recombination of neighbouring flux lines which brings a component of the current density along the local magnetic field, $\textbf{J}_{\parallel}$. This component generates distortions which also become unstable when a threshold value $J_{c\parallel}$ is exceeded, giving place to the so-called flux cutting phenomenon. Thus, when the conditions $J_{\parallel}=J_{c\parallel}$ and $J_{\perp}=J_{c \perp}$ become active, the so-called {\em double critical} state appears.
%

%
From the mathematical point of view, the critical state problem consists of finding the equilibrium distribution for the circulating current density $\textbf{J}({\bf r})$ defined
by the conditions $J_{\parallel}\leq J_{c\parallel}\; and \; J_{\perp}\leq J_{c\perp}$ both consistent with the Maxwell equations in quasistationary form, i.e.: displacement currents are neglected \cite{Ruiz_2009}.
Customarily, one also considers situations where the local components of the magnetic field $\textbf{H}(\textbf{r})$ along the superconductor are much higher than the lower critical field $H_{c1}$ and well below $H_{c2}$ to allow the use of the linear relation $\textbf{B}=\mu_{0}\textbf{H}$.

%
%
The general statement of the critical state, in the above described terms, was done by Clem and P\'erez-Gonz\'alez~ \cite{DCSM_I,DCSM_II,DCSM_III,DCSM_IV,DCSM_V,DCSM_VI,DCSM_VII,DCSM_VIII}. In particular, these authors have provided the physical background for successfully understanding an important number of experiments with rotating and oscillating magnetic field components. 
On the other hand, the theoretical scenario has been successively enlarged by a number of alternative approaches that focus on different aspects of the vast number of experimental activities in this field, e.g., one can identify the so-called:
\begin{enumerate}
\item {\em Isotropic models}\cite{Isotropic_I,Isotropic_II,Isotropic_III,Isotropic_IV,Isotropic_V} in which the critical state hypothesis reads $J_{\parallel}^2+J_{\perp}^2 \leq J_{c}^2$ in the spirit of Bean's original hypothesis \cite{Bean_1970}.
\item The {\em Elliptical model}\cite{Elliptical_I,Elliptical_II,Elliptical_III}, posed through the condition $J_{\parallel}^2/J_{c\parallel}^2+J_{\perp}^2/J_{c\perp}^2 \leq 1$.
\item The so-called {\em T-states} characterized by $J_{\perp}\leq J_{c\perp}\; and \;J_{\parallel}$ unbounded \cite{Brandt_2007}.

\end{enumerate}
%

%
In this work, we will show that all the above mentioned models may be unified within a continuous two-parameter theory that poses the problem  in terms of geometrical concepts within the $J_{\parallel}-J_{\perp}$ plane. To be specific, in this framework, we show that by the application of our variational statement \cite{Ruiz_2009}, one is able to specify almost any critical state law by means of an integer index $n$, that accounts for the smoothness of the $J_{\parallel}(J_{\perp})$ relation, and a certain {\em bandwidth} characterising the magnetic anisotropy ratio $\chi\equiv J_{c\parallel}/J_{c\perp}$. This will allow to elucidate the relation between diverse physical processes and the actual material law.
%

\begin{figure}
\begin{center}
{\includegraphics[width=0.48\textwidth]{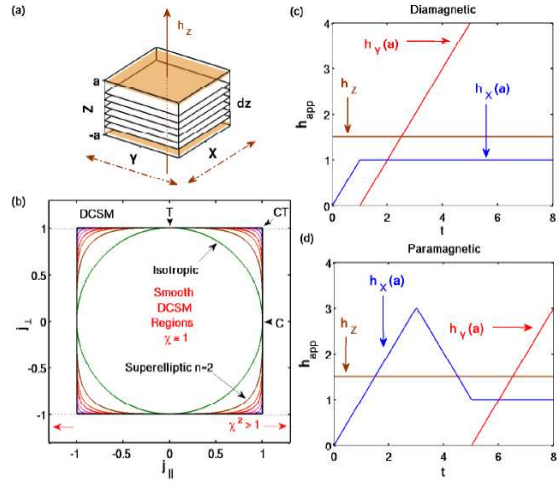}}
\caption{\label{Fig_1}(Color online) (a) Pictorial illustration of the slab geometry with a perpendicular magnetic component $h_{z}$. (b) Geometrical interpretation of the smooth double critical state model ({\em Sm}-DCSM) with $\chi=j_{c\parallel}/j_{c\perp}$=1. Several regions are incorporated through the order of the superelliptical functions ($n=1,2,3,4,6,10,20,40,\infty$). Here $n=1$ corresponds to the isotropic model (circular region), and $n\to\infty$ corresponds to the classical DCSM (square region) introduced by Clem and P\'{e}rez-Gonz\'{a}lez~\cite{DCSM_I}. (c) Schematics of the time dependence of the applied magnetic fields in the diamagnetic configuration, and (d) in the paramagnetic configuration.}
\end{center}
\end{figure}

%
The paper is organised as follows. In Sec.~\ref{Section_II}  we put forward some details about the most remarkable features of the critical state theory. The physical interpretation of the underlying approximations is focused on within our variational formulation. Then, Sec.~\ref{Section_III} is devoted to observe some properties of the electrodynamical behavior of the superconductor for different choices of the parameters $n$ and $\chi$. Specifically, we consider 3D magnetic field configurations in the infinite slab geometry for different initial states and processes (see Fig.~\ref{Fig_1}). A global discussion of our results and some concluding remarks are finally presented in Sec.~\ref{Section_IV}.

%
%

%


\section{Magnetic anisotropy of the Critical State}~\label{Section_II}
%

%
As stated above, the most complete description of irreversible phenomena in type-II superconductors at a macroscopic level is done through the commonly called Double Critical State Model (DCSM) introduced by Clem and P\'erez-Gonz\'alez \cite{DCSM_I,DCSM_II,DCSM_III,DCSM_IV,DCSM_V,DCSM_VI,DCSM_VII,DCSM_VIII}. Let us recall some basic ideas that will be essential for our further treatment. The material law introduced by the above authors in the form of the threshold conditions $J_{\parallel}\leq J_{c\parallel}$ and $J_{\perp}\leq J_{c\perp}$ for the current density flowing either parallel or perpendicular to the local magnetic field has been expressed in a geometrical language in previous articles \cite{Badia_2002,Ruiz_2009}. Essentially, our concept is to define a region $\Delta_{\textbf{r}}(\textbf{J})$ within the $J_{\parallel}-J_{\perp}$ plane, such that nondissipative current flow occurs when the condition $\textbf{J}=\textbf{J}_{\parallel}+\textbf{J}_{\perp}\in\Delta_{\textbf{r}}$ is verified. On the contrary, a very high dissipation is to be assumed when $\textbf{J}$ is driven outside $\Delta_{\textbf{r}}$. Recall that this scheme has allowed to translate the DCSM physics onto a region of currents defined in 3D by a cylinder \cite{Ruiz_2009} with its axis parallel to the local magnetic field $\textbf{B}$, and a rectangular longitudinal section in the plane defined by the vectors $\textbf{\textbf{J}}_{\parallel}=J_{c \parallel}\textbf{\^{E}}_{\parallel}$ and $\textbf{J}_{\perp}=J_{c \perp}\textbf{\^{E}}_{\perp}$. Such section is shown in Fig.~\ref{Fig_1}. We recall that in 2D problems with in-plane currents and magnetic field, the current density region straightforwardly coincides with the above mentioned longitudinal section.
%
%
In this scheme, the parts of the sample where only the flux depinning threshold has been reached are denoted T zones or flux transport zones ($J_{\perp} = J_{c \perp}~;~J_{\parallel}<J_{c \parallel}$). They are represented by points in a horizontal band. Physically, the flux lines are migrating while basically retaining their orientation. On the other hand, regions where only the cutting threshold is active are denoted as C zones or flux cutting zones ($J_{\parallel} = J_{c \parallel}~;~J_{\perp}<J_{c \perp}$). They are represented by points in a vertical band. In those regions where both mechanisms have reached their critical values are defined as CT zones ($J_{\parallel} = J_{c \parallel}$ and $J_{\perp} = J_{c \perp}$). The current density vector belongs to the corners of a rectangle. Finally, the regions without energy dissipation are called O zones, and the current density vector belongs to the interior of the rectangle. 

%
In this section, and corresponding to the regions depicted in the lower part of Fig.~\ref{Fig_1}, we investigate the magnetic response of type-II superconductors, whose material law is obtained by smoothly modifying the standard DCSM rectangular region until the elliptic cases are reached. In this work, we focus on the role of the smoothing index $n$, and thus, will only consider the extreme cases $\chi =1$ and $\chi\to\infty$. Notice that the smooth limiting cases (elliptic or isotropic) can be considered as the manifestation of {\em averaged} values of the critical current restrictions due to the inhomogeneity of the material, as well as a consequence of flux line interactions at a mesoscopic level that introduce coupling between the thresholds $J_{c \parallel}$ and $J_{c \perp}$. In any case, smooth models have to be considered as related to a number of experiments that one could not explain within {\em piecewise} smooth statements \cite{Voloshin_1997,Fisher_1997,Fisher_2000,Voloshin_2001,Fisher_2000_b,Voloshin_2010}. Hereafter, we will use the notation {\em Sm}-DCSM for such models.

\subsection{Variational Statement}

As indicated above, the selection of appropriate restrictions for the macroscopic current density is a significant step forward to reveal the intrinsic structure of the mechanisms involved. In conventional approaches, related to the material law ${\bf J}({\bf E})$, and starting from the Maxwell equations, one may obtain the penetration profiles for the magnetic field from a differential equation statement of diffusive type (i.e.: $\partial_{t}{\bf H}=f(\nabla^{2}{\bf H})$). In our case, the basis of the \textit{Sm}-DCSM relies in a parallel (and equivalent) formulation that uses a discretisation scheme of the magnetic field in terms of time-steps connected by the finite-difference expression $\mu_{0}(\textbf{h}_{l+1}-\textbf{h}_{l})$. The evolution from one magnetostatic state to another is obtained variationally. Thus, we minimise the functional 
%
%
\begin{equation}\label{Eq_1}
\textit{F}[\textbf{h}_{l+1}(\textbf{r})]=\frac{\mu_{0}}{2}\int_{\forall} |\textbf{h}_{l+1}-\textbf{h}_{l}|^{2}+\textbf{p}\cdot(\nabla\times \textbf{h}_{l+1}-\textbf{j})
\end{equation}
where the Lagrange multiplier enforces Amper\`{e}'s law~\cite{Badia_2001,Ruiz_2009}. In addition, the minimization is performed with the local distribution of currents constrained by the law ${\bf J}\in \Delta_{\textbf{r}}$. Notice that, either material or extrinsic anisotropy can be easily incorporated by prescribing $\Delta_{\textbf{r}}$ to be the appropriate region. For instance, by modeling $\Delta_{\textbf{r}}$ as an elliptical~\cite{Elliptical_I,Elliptical_II,Elliptical_III,Badia_2002,Ruiz_2009} or a rectangular~\cite{DCSM_I,DCSM_II,DCSM_III,DCSM_IV,DCSM_V,DCSM_VI,DCSM_VII,DCSM_VIII,DCSM_IX,Ruiz_2009} region oriented over selected axes. Mathematically, such kind of regions are hosted as limiting cases of a smooth expression defined by the two-parameter family of superelliptic functions
%
%
\begin{equation}\label{Eq_2}
\left(\frac{j_{\parallel}}{j_{c \parallel}}\right)^{2n}+\left(\frac{j_{\perp}}{j_{c \perp}}\right)^{2n}\leq 1.
\end{equation}
%

%
The reader can immediately verify that an index $n=1$ and a bandwidth defined by $\chi\equiv j_{c \parallel}/j_{c \perp}=1$ correspond to the standard isotropic model~\cite{Isotropic_I}. 
On the other hand, when one assumes enlarged bandwidth (i.e.: $\chi >1$), the \textit{Sm}-DCSM becomes the standard elliptical model introduced by Romero-Salazar and P\'erez-Rodr\'iguez~\cite{Elliptical_I,Elliptical_II}. 
When the bandwidth $\chi$ is extremely large, i.e., $J_{c\parallel}\gg J_{c \perp}$, one recovers the so-called $T-$states treated by Brandt and Mikitik\cite{Brandt_2007}. Rectangular regions strictly corresponding to the DCSM\cite{DCSM_I,DCSM_II,DCSM_III,DCSM_IV,DCSM_V,DCSM_VI,DCSM_VII,DCSM_VIII} are obtained for the limit $n\rightarrow\infty$ and arbitrary $\chi$. Finally, allowing $n$ to take values over the positive integers, a wide scenario describing anisotropy effects is envisioned [Fig.~\ref{Fig_1}(b)]. Such regions will be named after superelliptical and their properties can be understood in terms of the rounding (or smoothing) of the corners for the DCSM.
\subsection{Numerical treatment}

%
In order to illustrate the effect of the material law, below we will show the behavior of the field and current-density profiles for the system illustrated in Fig.~\ref{Fig_1} with different selections of the region $\Delta_{\textbf{r}}$.
To be specific, we have considered an infinite slab made up by $2\cdot N_{S}$ longitudinal sheets arranged along the x-y plane and filling the space $|z|\leq a$. Symmetry or antisymmetry conditions for the different electrodynamical quantities can be be applied along the $N_{s}$ sheets arranged in $0\leq z\leq a$. On the other hand, one can assume in-plane position independence of {\bf J}, i.e.: one has $[J_{x}(z_{i}),J_{y}(z_{i})]$. Note, in passing, that this ensures a divergenceless current density as required by charge conservation in steady states. Incidentally, we have to mention that along this work we have used $N_{s}=300$. 
From the technical point of view, we also mention that the physical parameters that define the problem have been used to renormalise the physical quantities. Thus, we use $\textbf{h}\equiv \textbf{H}/J_{c \perp}a$, and $\textbf{J}\equiv \textbf{J}/J_{c \perp}$. Recall that, one may assume the numerical value $J_{c \perp}$ as known {\em a priori} or obtained from experiment. Finally, the position within slab will be expressed in terms of ${\rm z}\equiv z/a$.
%

%
Now, following Ref.\cite{Ruiz_2009} the variational statement for the \textit{Sm}-DCSM, in numerical form leads to minimise the function
%
%
\begin{eqnarray}\label{Eq_3}
\textit{F}[\{ I_{i,l+1}\}]=&& \frac{1}{2}\sum_{i,j}I_{i,l+1}M_{ij}I_{j,l+1}-\sum_{i,j}I_{i,l}M_{i,j}I_{j,l+1}\nonumber \\ &&+\sum_{i}I_{i,l+1}\Delta h({\rm z}_{i})\, ,
\end{eqnarray}
where the set of unknown current values for the time layer $l+1$, i.e.:$\{I_{i,l+1}\}$ are defined within a collection of circuits (indexed by {\em i, j}) whose mutual inductance coefficients are represented by $M_{i,j}$. In the slab symmetry the circuits are just sheets made up of straight lines along the $x$ and $y$ axes. Finally, $\Delta h({\rm z}_{i})$ defines the time discretisation of the applied magnetic field ($\Delta h({\rm z}_{i})\equiv {h}_{y,l+1}-{h}_{y,l}$). As it was shown in Ref.\cite{Ruiz_2009}, the geometrical coefficients $M_{i,j}$ are given by
%
%
\begin{eqnarray}\label{Eq_4}
M_{i,j}^{x,y}\equiv&& 1+2[min\{i,j\}]_{i\neq j}\, ,~or \nonumber \\
\equiv&& 2\left(\frac{1}{4}+i-1\right)_{i=j~.}
\end{eqnarray}
Recall that inductive coupling only occurs for $x$ and $y$ layers separately, and that the corresponding coefficients are identical.

Eventually, the response of the superconductor is obtained as a pair of surface current functions $\{j_{x}({\rm z}_{i}),j_{y}({\rm z}_{i})\}$ for each one of the $N_{s}$ sheets. The magnetic field profiles and magnetic moments may be obtained by numerical integration. Thus, the magnetic moment components per unit area are obtained from
%
%
\begin{eqnarray}\label{Eq_5}
\textbf{M}=\int_{-a}^{a}\textbf{z}\times \textbf{j}~dz
\end{eqnarray}
%


\section{Results}~\label{Section_III}
%

%
\begin{figure*}
\begin{center}
{\includegraphics[width=0.98\textwidth]{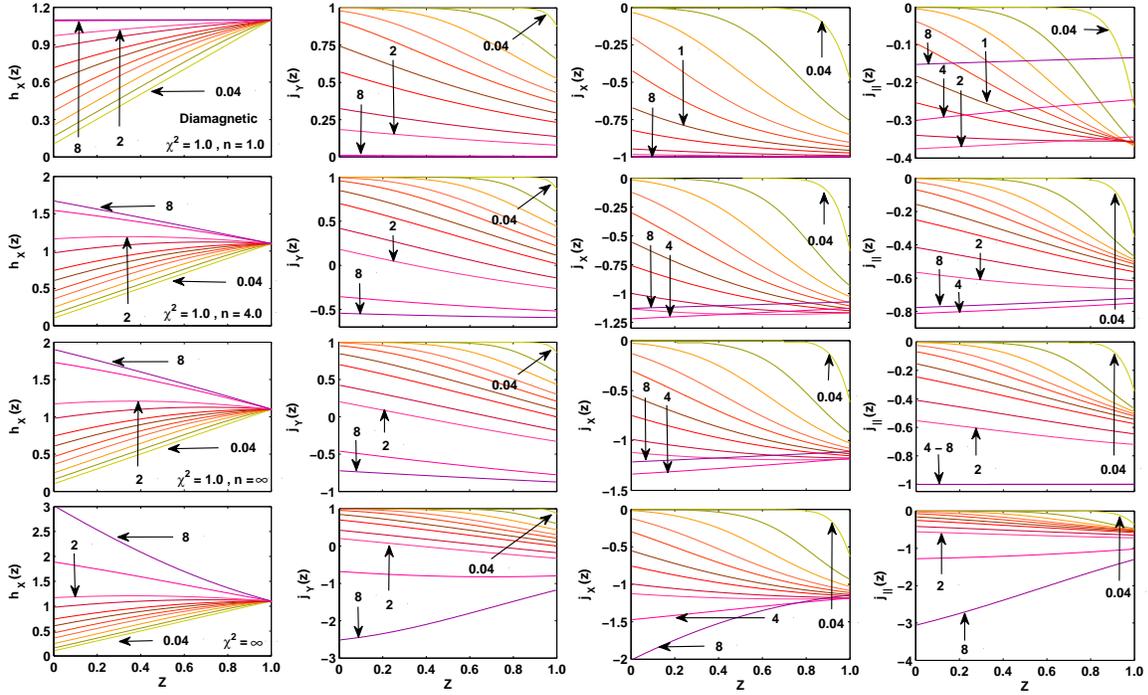}}
\caption{\label{Fig_2}(Color online) Profiles of the magnetic field component $h_{x}[z,h_{y}(a)]$ and their corresponding current-density profiles $j_{y}[z,h_{y}(a)]$ starting from the \textit{first}-time step defined by $h_{x}(a)=1.1$ and $h_{z}=1.5$ in the diamagnetic configuration (Fig.~\ref{Fig_1}(c)). The current component $j_{x}[z,h_{y}(a)]$ and the cutting component $j_{\parallel}$ are also shown. The curves are labelled according to the longitudinal magnetic field component at surface of slab corresponding to the values $h_{y}(a)=0.040, 0.2, 0.4, 0.6, 0.8, 1.0, 1.2, 1.6, 2.0, 4.0, 8.0$, and $h_{z0}=1.5$. First row: profiles for the isotropic model $\chi^{2}=1$, $n=1$. Second one: profiles for the \textit{Sm}-DCSM with $\chi^{2}=1$, $n=4$. Third one: profiles for the DCSM with $\chi^{2}=1$, $n=\infty$. Finally, in the lowest row, the profiles for the \textit{T-}state model, i.e., $\chi^{2}\to\infty$, $\j_{c\perp}\neq 0$ are shown.}
\end{center}
\end{figure*}
\begin{figure*}
\begin{center}
{\includegraphics[width=0.98\textwidth]{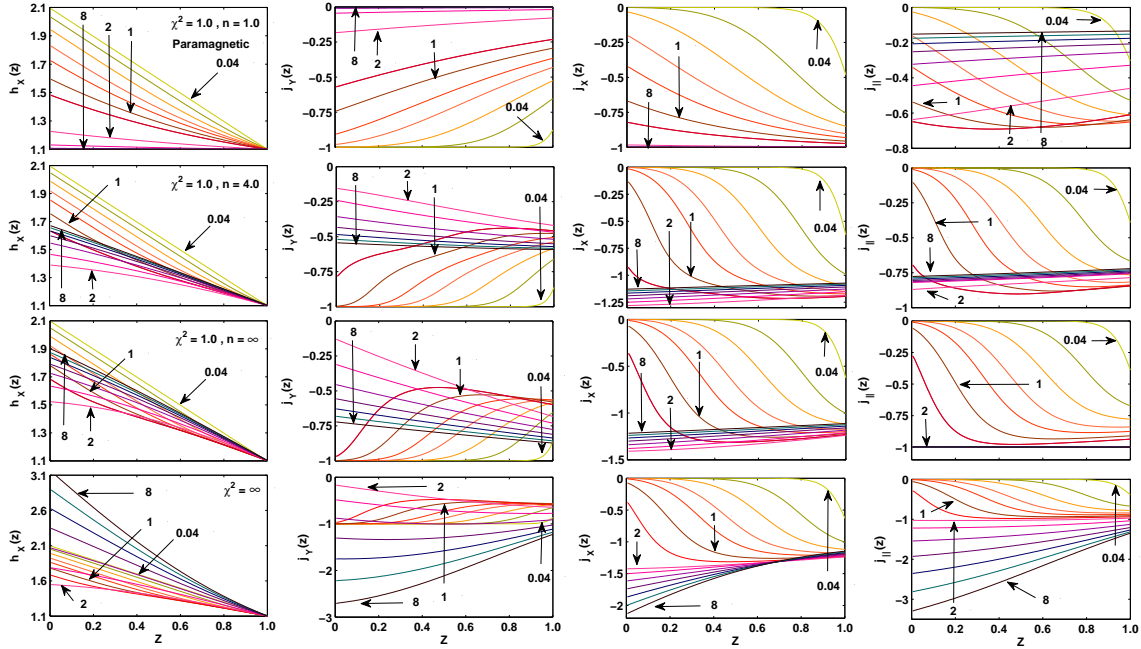}}
\caption{\label{Fig_3}(Color online) Same as Fig.~\ref{Fig_2}, but the initial paramagnetic configuration illustrated in Fig.~\ref{Fig_1}(d) with $h_{x}(a)=1.1$ and $h_{z}=1.5$. Here, the curves have been labelled according to the values $h_{y}(a)=0.040, 0.2, 0.4, 0.6, 0.8, 1.0, 1.2, 2.0, 3.0, 4.0, 5.0, 6.0, 7.0, 8.0$.}
\end{center}
\end{figure*}

%
%
%
Below, we present the theoretical predictions for the \textit{Sm}-DCSM with several choices for the index $n$ along the magnetization processes indicated in Fig.~\ref{Fig_1}. First, we consider solutions for ${\bf h}_{l+1}$, starting from a fully penetrated state with a magnetic field perpendicular to the slab surfaces, i.e., a lattice of parallel vortices is assumed to nucleate parallel to the $z$-axis within the sample. Then, the material is subjected to a surface field $h_{x}(a)$. As the external magnetic field is now augmented by means of $\Delta h_{x}(a)$, flux lines are tilted and penetrate the specimen until an equilibrium distribution is achieved (diamagnetic). If the external magnetic field is subsequently lowered, thereby reducing the retaining magnetic pressure, flux lines migrate out of the sample until the equilibrium is restored (paramagnetic). These initial configurations can be seen in Figs.~\ref{Fig_2}~\&~\ref{Fig_3}. From this point, the sample is immersed in a growing applied magnetic field along the y-axis inducing an additional inclination of the flux lines. 
%

%
%
%
%

\begin{figure}
\begin{center}
{\includegraphics[width=0.48\textwidth]{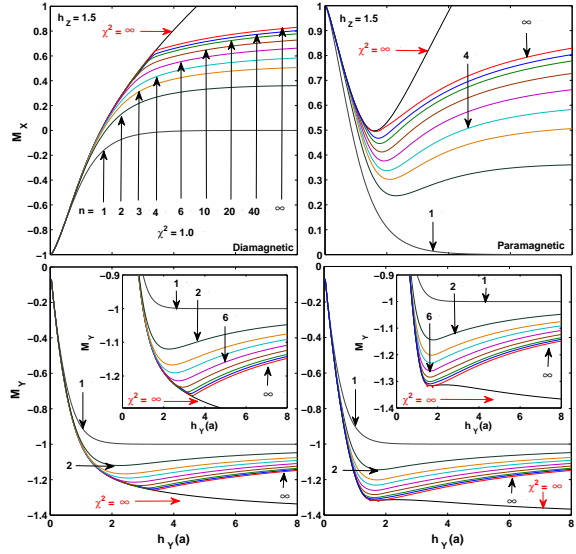}}
\caption{\label{Fig_4}(Color online) The magnetic moment components $M_{x}$ (top) and $M_{y}$ (bottom) per unit area as a function of the applied magnetic field component $h_{y}(a)$ in the diamagnetic (left) and paramagnetic (right) initial configurations defined by $h_{x}(a)=1.1$ and $h_{z}=1.5$. Results for several models are shown, e.g., the \textit{T-}state model ($\chi^{2}\to\infty$, $j_{c\perp}\neq 0$), the conventional DCSM ($n\to\infty$), the \textit{Sm}-DCSM with $\chi^{2}=1$ and $n=1$, and the \textit{Sm}-DCSMs with $\chi^{2}=1$ and $n=2, 3, 4, 6, 10, 20, 40$. The insets shows a zoom of the behavior around the minima.}
\end{center}
\end{figure}

%
In order to allow a physical interpretation on how the material law affects the response of the sample, in Figs.~\ref{Fig_2}~\&~\ref{Fig_3} we plot the profiles of magnetic field and induced currents as $h_{y}(a)$ is increased. In particular, we show the electrodynamic evolution of the critical state for regions defined by $\chi^{2}=1$ and indexes $n=1$ (isotropic model), $n=4$ (\textit{Sm}-DCSM with corner rounded), and $n\rightarrow\infty$ (rectangular DCSM), with the initial condition $h_{z}=1.5$ and $h_{x}(a)=1.1$ for both diamagnetic and paramagnetic states. In addition, the profiles for the condition $\chi^{2}\to\infty$ and $j_{c\perp}\neq 0$ (infinite bandwidth model or \textit{T}-state model) are also included. The magnetic response for other regions $\Delta_{\textbf{r}}$ with corners smoothed by the index condition $n=2,3,6,10,20,40$ are shown in Fig.~\ref{Fig_4}.
%

%
When the magnetic field $h_{y}(a)$ is switched on, an electric field $E_{x}$ arises in the surface layer of the superconductor according to the Faraday's law. This electric field produces a current $j_{x}$ that will screen the excitation. Then, owing to the restrictions on the current density vector ${\bf j}$ introduced by the material law, the local component $h_{x}(z)$ is pushed towards the center of the sample in the diamagnetic case, or towards the external surface in the paramagnetic one as one can observe in Figs.~\ref{Fig_2}~\&~\ref{Fig_3}. The local variation of the magnetic field $h_{x}$ relates to an inner variation of the current density along the y-axis ($j_{y}(z)$) which engenders flux cutting as related to a density of current flowing parallel at the flux-lines ($j_{c \parallel}$).
%

%
%
%
%
We recall the following features for the different critical current models that have been considered. Within the isotropic case (upper row of Figs.~\ref{Fig_2}~\&~\ref{Fig_3}), the local component $h_{x}(z)$ increases (diamagnetic case) or it reduces (paramagnetic case) until the specimen is fully penetrated to satisfy the condition $h_{x}= h_{x}(a)\;\forall~z$, i.e., $j_{y}(z)\rightarrow0$ as $h_{y}$ increases. As a consequence, no sign reversal in the induced currents is predicted. Thus, notice that the isotropic hypothesis ($\chi^{2}=1$, $n=1$) provides a straightforward explanation of the observed magnetization collapse~\cite{Isotropic_II,Isotropic_III,Isotropic_IV}.
Another related phenomenon, the so-called {\em paramagnetic peak effect} of the magnetic moment is not foreseen by the isotropic material law. In previous works, this observation was explored in terms of the so-called two velocity electrodynamic model as a crude approximation for the real dynamics in a flux line lattice~\cite{Voloshin_1997,Voloshin_2001,Fisher_1997,Fisher_2000_b,Voloshin_2010}. Here (Fig.~\ref{Fig_4}) we notice a pronounced peak effect in both components of the magnetic moment. We emphasise that whatever region is considered [excepting the limiting cases ``$\chi^{2}=1$, $n=1$'' (isotropic model), and ``$\chi^{2}\to\infty$'' (T- or infinite bandwidth- model)], the peak effect in the paramagnetic case is predicted for both components of the magnetization. In this sense, we argue that the peak effect cannot be interpreted as a direct evidence of an elliptical material law. Instead of this, it is a universal signal of the anisotropy effects involved in a general description of the material law. The evolution of the peak effect as a function of $\chi^{2}$ has been shown in Figs. 17 and 18 of Ref.~\cite{Ruiz_2009}. There, we note that an increase of the bandwidth $\chi^2$ produces a stretched magnetic peak. Consequently, paramagnetic effects are visible over a wider range as the cutting threshold value $j_{c \parallel}$ increases. We emphasise that the overall effect of increasing the value $\chi^{2}=(j_{c \parallel}/j_{c \perp})^{2}$ is that the components of ${\bf M}$ get closer to the {\em master} curves defined by $\chi\to\infty$. Several further distinctive signals for the different models are highlighted below.
%

%
On the one hand, for the isotropic model, the collapse of the magnetization is achieved while $j_{\parallel}$ is monotonically reduced (upper rows in Figs.~\ref{Fig_2} and \ref{Fig_3}). When the material law is the infinite bandwidth model or T-state model ($\chi^{2}=\infty$), the magnetization collapse does not take place, and there is no restriction on the longitudinal component of the current that increases arbitrarily towards the center of the sample. This corresponds to the absence of flux cutting, i.e.: $j_{\parallel}$ does not saturate by reaching a threshold value $j_{c \parallel}$. For rectangular or {\em smooth rectangular} regions (intermediate rows in Figs.~\ref{Fig_2} and \ref{Fig_3}), together with the absence of collapse, one also observes that $j_{\parallel}$ basically saturates to a value that depends on the smoothing parameter $n$ (exactly to $j_{c\parallel}$ for the very rectangular case $n\to\infty$).
%

%
Remarkably, when a rectangular section is assumed, the sample globally reaches the CT state (corner of the rectangle). As a consequence of the sharp limitation for $j_{\parallel}$, a well-defined corner in the magnetic moment components $M_{x}$ and $M_{y}$ appears, both for the diamagnetic and paramagnetic cases (see Fig~\ref{Fig_4}). This clear trace of the DCSM establishes the departure from the master curves defined by the \textit{T}-state, and has been assigned to the instant at which the sample reaches the CT state. We call the readers' attention about the noticeable gap in Fig.~\ref{Fig_4}, separating the isotropic model ($\chi^{2}=1,~n=1$) and the square model (``$\chi^{2}=1,~n\to\infty$''). Thus, the question arises about the possibility of {\em inverting} an experimental response within this interval so as to uniquely determine a given smooth region $\Delta_{\textbf{r}}$ for the components of $\bf J$. As it will be argued below, complimentary information about the limitations $J_{c\parallel}$ and $J_{c\perp}$ is due. Thus, if one compares Fig.~\ref{Fig_4} and Figs. 17 and 18 in Ref.\cite{Ruiz_2009} one can realize that smooth models for a given ratio $\chi\equiv J_{c\parallel}/J_{c\perp}$ will fill the gap between the master limiting curves defined by the rectangular ($\chi , n\to\infty$) and elliptic ($\chi\, , n=1$) models, and their corresponding curves for different values of $\chi$ will intersect in a complicated fashion. In other words, the magnetization curves by themselves do not provide an exhaustive information on the material law which defines the critical state dynamics in type II superconductors. In fact, notice that in the regime of low fields $h_{z}\sim h_{y}(a)$ (or a weak oscillating magnetic field in presence of a strong constant field~\cite{Isotropic_I,Isotropic_IV}) the material law is indistinguishable and the magnetic moment may be reproduced even by the isotropic model.
%

%
Nevertheless, a deeper insight in Figs.~\ref{Fig_2} and \ref{Fig_3} reveals that the local behavior of the current density profiles, if available, should give clear indications. Thus, notice that although the dynamics of the profiles $h_{x}$, $j_{y}$, and $j_{x}$ is almost indistinguishable between the smooth and rectangular models (third and fourth rows of Figs.~\ref{Fig_2} and \ref{Fig_3}), a clear distinction arises by analysing $j_{c \parallel}$. On the one hand, when the rectangular model is assumed $j_{\parallel}$ reaches the threshold value $j_{c \parallel}$, and the entire specimen verifies a CT-state as the applied magnetic field increases. On the other hand, when the rectangular region is smoothed by the index $n$, the parallel component of the current density eventually decreases to a value that depends on the values of $n$ and $\chi$.
%
%


\section{Conclusions}~\label{Section_IV}

In type-II superconductors, an incomplete isotropy for the limitations of the current density relative to the orientation of the local magnetic field arises from the different physical conditions of current flow either along or across the Abrikosov vortices. One can thus talk about magnetically induced anisotropy. In this work, we have explored the application of the so-called {\em smooth double critical state model} to anisotropic material laws in such sense. Theoretical predictions have been made that allow to establish a relation between a number of experimental observations and anisotropies of the material law. Two fundamental material-dependent quantities play key roles in this theory (${J}_{c\parallel}$,${J}_{c\perp}$) related to the flux cutting and depinning thresholds. We have applied the theory to the case of a type-II superconducting slab assuming translational symmetry along its surface and a 3D magnetic field.

Motivated by the possibility of modelling the influence and mutual interaction between the critical current thresholds, we have investigated situations with current density vectors belonging to some smooth region $\Delta_{\textbf{r}}$ within the ${J}_{\parallel}-{J}_{\perp}$ plane.
This has been done by describing the boundary of $\Delta_{\textbf{r}}$ by means of a superelliptic relation ($({J}_{\parallel}/{J}_{c\parallel})^{2n}+({J}_{\perp}/{J}_{c\perp})^{2n}= 1$). Notoriously, the material law $\Delta_{\textbf{r}}$ is determined by the index $n$ and a proper {\em bandwidth} $\chi\equiv J_{c \parallel}/J_{c \perp}$. Thus, our predictions cover a wide range of laws: (i) the isotropic model ($\chi^{2}=1$, $n=1\Rightarrow \Delta_{\textbf{r}}$ is a circle), (ii) the elliptical model ($\chi^{2}>1$, $n=1\Rightarrow \Delta_{\textbf{r}}$ is an ellipse), (iii) the rectangular model ($\chi^{2}\geq 1$, $n\to\infty\Rightarrow \Delta_{\textbf{r}}$ is a rectangle), and (iv) the infinite band model ($\chi^{2}\rightarrow\infty$). After a detailed analysis that entails the local electrodynamics for material laws that cover a wide range of values for $n$ and $\chi$ we conclude that a considerable amount of experimental observations may be explained in this framework. Thus, we have shown that: (i) the magnetic moment collapse by a perpendicular field is clearly assigned to the isotropic behavior of $\bf{J}_{c}$, (ii) paramagnetic magnetization induced by the application of a perpendicular field is always predicted if anisotropy in the region $\Delta_{\textbf{r}}$ is allowed. (iii) Paramagnetic peak effects induced by a perpendicular magnetic are expected in a wide range of conditions, (iv) differences between the several models studied are smeared out for low magnetic fields, (v) unless for the extreme cases (isotropic $\chi =1 , n=1$ and T-states $\chi\to\infty$) the {\em inversion} of magnetic data ($M_{x},M_{y}$) so as to elucidate the specific critical state region for a given sample is not straightforward. Complimentary information about the maximal values of ${J}_{c\parallel}$ and ${J}_{c\perp}$ is required so as to extract the complete material law ${J}_{c\parallel}({J}_{c\perp})$. Further research along this line is suggested, i.e.: the design of some experimental routine that defines a well posed inverse problem for the determination of $\Delta_{\textbf{r}}$.


\section*{Acknowledgements}

We would like to thank Dr. C. L\'{o}pez for a number of useful discussions. This work was supported by the Spanish CICyT project MAT2008-05983-C03-01. H. S. Ruiz acknowledges a grant from the Spanish CSIC-JAE program.

\vspace{0.5 cm}

%
%

\end{document}